# ADAPTATION OF WEB SERVICES TO THE CONTEXT BASED ON WORKFLOW: APPROACH FOR SELF-ADAPTATION OF SERVICE-ORIENTED ARCHITECTURES TO THE CONTEXT

Faîçal Felhi[1] and Jalel Akaichi[2]

[1, 2] Department of Computer Sciences, ISG, BESTMOD,Tunis, Tunisia
[1]felhi_fayssal@yahoo.fr , [2] jalel.akaichi@isg.rnu.tn

## ABSTRACT

*The emergence of Web services in the information space, as well as the advanced technology of SOA, give tremendous opportunities for users in an ambient space or distant, empowerment and organizations in various fields application, such as geolocation, E-learning, healthcare, digital government, etc.. In fact, Web services are a solution for the integration of distributed information systems, autonomous, heterogeneous and self-adaptable to the context. However, as Web services can evolve in a dynamic environment in a well-defined context and according to events automatically, such as time, temperature, location, authentication, etc.. We are interested in improving their SOA to empower the Web services to be self adaptive contexts. In this paper, we propose a new trend of self adaptability of Web services context. Then applying these requirements in the architecture of the platform of adaptability to context "WComp", by integrating the workflow. Our work is illustrated by a case study of authentication.*

## 1. INTRODUCTION

The evolution of information systems as well as pervasive systems is designed to make information available anywhere and anytime. For this case, considerable approaches related to adaptability with different modes of implementation such as: AOP (Aspect Oriented Programming)[8]. This aspect used by various platforms on the goal to adapt the Web service to the context dynamic changes of environment. The emergence of Web services as a model for integrating heterogeneous web information has opened up new possibilities of interaction and adaptability to context when offered more potential for interoperability. However, from a set of requirements on SOA, and to provide self adaptation to the context of web services, we need to integrate more generic connector that takes into account all ambient or distant events.

Based on technology platforms to adapt Web services context, and also gain benefit from the advantages of Web services. These platforms deal with "simple" adaptation to functional and technical exchange purpose it is by no means clustering adaptation to ambient context year. These systems must be used in different contexts depending on the environment of the user profile and the terminal to use. One of the major problems of such systems therefore relates to the context adaptation.

Our goal is to study the adaptability with a presentation of platforms to adapt their operating principles, and trying to improve the SOA to empower the Web services to be self adaptive to the contexts. For this raison, We used WComp platform [13][14][15] as a middleware for context adaptation to show the feasibility of our approach, and illustrated our work by a case study.

## 2. BASIC CONCEPTS
### 2.1. Web services

Web services (WS) [27], like any other middleware technologies, aim to provide mechanisms to bridge heterogeneous platforms, allowing data to flow across various programs. The WS technology looks very similar to what most middleware technologies looks like. Consequently, each WS has an Interface Definition Language, namely WSDL (Web Service Description Language) [28], that is responsible for the message payload, itself described with the equally famous protocol SOAP (Object Access Protocol) [29], while data structures are explained by XML (eXtended Markup Language) [30]. Very often, WS are stored in UDDI (Universal Description Discovery and Integration) registry [31].

In fact, the winning card of this technology is not its mechanism but rather the standards upon which it is built. Indeed, each of these standards is not only open to everyone but, since all of them are based on XML, it is pretty easy to implement these standards for most platforms and languages. For this reason, WS are highly interoperable and do not rely on the underlying platform they are built on, unlike many ORPC (Object Request Procedure Call). Web services standards are gathered in WSA (Web Service Architecture) [27][32].

### 2.1. Adaptation

It is important that applications adapt to their surroundings [22]. The adaptation software can take many forms, and we refer to "system adaptable "when a user can interact with the system and, through them, modify and customize. An adaptive system identifies a situation and adapts. Activation of this system changes can be caused by human intervention or a number of observations.

### 2.2. Adaptability, self-adaptability and adaptivity

If the user has the possibility to adapt the interaction of these preferences, the interface is said to be configurable and adaptable [12].

The adaptability of a system is the software mechanisms that achieve system changes. It is these mechanisms that modify the behavior of the system while preserving the properties of the system. Strategies to control the adaptivity mechanisms of adaptability. The self-adaptability means the power to dynamically modify the behavior of a system in response to internal and external events. If the system is able to adapt his behavior to the needs (capabilities and preferences of the current user) during the interaction, with its capacities of perception and interpretation of the interaction and its context, the interface is called adaptive. The adaptivity of a system is how the system adapts. It consists of strategies to trigger changes in the system based on incentives.

### 2.3. Ambient computing, ubiquitous computing, and cloud computing

The ambient computing [21] around us. It is invisible. This does not mean that you can not see with our eyes. This applies to that on which our attention and awareness focus. Thus, technologies tend to disappear to mix with elements of everyday life.

The ambient computing also known as ambient intelligence refers to environments sensitive to the physical world. Ubiquitous computing goes beyond computerized real objects, it blends with the environment every day soaking in the physical phenomena that are manifested.

Cloud computing [32][19][9] is a model for enabling convenient, "on-demand" network access to a shared pool of configurable computing resources (e.g., networks, servers, storage, applications, and services) that can be rapidly provisioned and released with minimal management effort or service provider interaction. This cloud model promotes availability and is composed of five essential characteristics, three service models, and four deployment models as: On demand self-service, Ubiquitous network access, Resource pooling, Location independence, Homogeneity, Rapid elasticity, Measured service. Cloud computing brings a new level of efficiency and economy to delivering IT resources on demand. It offers efficiency and agility.

## 2.4. Context and ambient context

Etymologically, the word context, which has its origins in "co-text" is defined as the entire text surrounding a drawn and which illuminates its meaning. Over time its meaning has evolved to be defined as follows: "The set of circumstances in which an event occurs, is an action" [33].
The notion of context is used in many fields such as artificial intelligence, linguistics, psychology, and of course, ambient computing.

The main goal of context execution is to allow the application to manage the external situations that affect its quality of service seen by the user. Therefore, the application must be adaptable to disappearances and appearances of devices on the network, for example, or all sorts of technical failures, malicious or unusual expense.

The context is not directly involved in the application, but it sets the execution environment of the application that is a subset of the software infrastructure. It is in fact available resources of the system at a given [25].

## 2.5. Work Flow

WWF (Windows Workflow Foundation) [40][41] is a framework that allows users to create flow systems or user applications written for Windows Vista, Windows XP and Windows Server 2003 family. WWF can solve simple scenarios such as displaying user interface controls based on user input or complex scenarios encountered by large companies, such as order processing and inventory control.

WWF treat the flux activation in business applications, the flow of pages the user interface, the flow of paper, user flows, mixed flows for applications based on services, and flow-driven rules business [39].

## 3. RELATED WORKS
### 3.1. Aspect-based Approaches

Many approaches based on aspects.
AspectWerkz [10], supporting weaving at loading, offers various means of expression of cut points. It offers different methods to specialize or customize the middleware according to the code on the works. Furthermore, it supports different various methods of expression (XML and Java).

DynamicTAO [6] have a particularity that can represent different aspects of the ORB [2] in the form of strategy design pattern. The system configuration is offer via a file (specifies the different strategies applied by the ORB before launching the system). Based on our work, we conclude that this platform suffers like all conventional middleware for the inability to reconfigure the ORB during execution time.

DynamicTAO, address aspects such as: competition, safety and supervision, and can provide a set of interfaces allowing users to reconfigure its structure.

## 3.2. Aspect and Web Service-based Approaches

Charfi and al. approach [1] propose a framework that provides support for middleware BPEL (Business Process Execution Language) [34] engines. The authors apply the concepts of deployment descriptor and container for the Web service composition.

Ferraz Tomaz and al. approach [23] proposed a tool for weaving aspects for a simple adaptability of the Web services, implementing aspects of the services as loosely coupled, where aspects are woven dynamically. In this approach, aspects are themselves Web services, thus they are independent of languages and platforms. This approach provides a mechanism, based on the AOP for Web services, to dynamically adapt to different policies of use.

This approach has two major limitations related to its architecture and its implementation. These limitations are: the dependence of the runtime architecture of Web services and dependence on aspects of language.

Mehdi Ben Hmida approach [17] extended the solution proposed by [23] to specify BPEL (Business Process Execution Language) [34] processes adaptable, that is to say, the adaptability of complex services. This specification allows the generation of customers that adapt to dynamic changes on the server side. Changes made to the service can lead to further exchanges of information between client and server that were not initially planned during implementation of the client application. These new interactions normally cause runtime errors of the customer. This approach is based on process algebras to dynamically generate customers. Process algebras manage the interactions between a BPEL process and its customers, this by specifying formally the interaction protocol (abstract BPEL) and automatically generating a client who is successfully communicating with the service. This approach overcomes the limitations in the dynamic modification of a process that can lead to a change in the pattern of interaction with the client and will fit the client and server parts. Hence the need to extend the semantic aspects and Web services, which resulted in the ASW (Aspect Service Weaver).

Aspects are themselves loosely coupled Web services, they are independent of languages and platforms, but, this approach has limitations. Adaptation to context is not taken into account, that is to say, if an event occurred during a search on a Web service, this approach does not take into account this event.

## 3.3. Context Adaptation-based Approaches

The ambient computing encourages the proliferation of associated devices. A key aspect of the ambient computing is its invisibility. Users perceive the features but do not see the devices that provide these features. Adaptability and evolution of software in these devices becomes an asset to their condition of use. There are various of ambient information systems available as Aura [5], ExORB [20], DoAmI [16], CORTEX [7], WComp [13][14][15].

From studies by [4] we can summarize the descriptions of the following platforms:

Aura [5] is a context-sensitive middleware that enables the design of mobile applications. Aura's goal is to provide each user with a set of implicit processing services and information.

ExORB demonstrates the ability of a middleware to support its configurability, the possibility of putting it-to-day improvement. ExORB examine in particular middleware for mobile phones.

Such devices require a middleware on which it is possible to configure new software, to improve the software already built without manual intervention by the end user. ExORB purposed of contributing to the construction of middleware services for strengthening the main features of configurability, the ability to update and improvement.

DoAmI (Domain-Specific Ambient Intelligence) deals with the dynamic aggregation of distributed services. Its offers an architectural model that relies on a service-oriented middleware for integration and activation of services at runtime.

CORTEX (Co-operating Real-time sentient objects: architecture and Experimental Evaluation) aims to build a middleware component-based applications to accommodate influenced by the external environment, particularly in the transportation field.

CORTEX needs and capabilities of local services for local decision making. The system participates in a cooperative global system. Thus, it has a system for collecting real-time environment.

In addition, local systems can be equipped with additional features such as consideration of traffic lights or a mechanism to allow pedestrians (presence sensor, obstacle avoidance).

CORTEX defines objects aware. They are moving objects that behave independently and are responsible for interactions with the physical environment. These behaviors are based on sensor inputs and the internal state of each object.

To address the problems of coordination, control, adaptability and scalability, CORTEX provides a first programming paradigm using objects aware.

The discovery mechanism is not well detailed by the authors and does not measure the scalability of the architecture. In addition, it lacks the tools to do a self adaptation.

WComp represent the implementation of experimental models presented in the work of RAINBOW (research team of the I3S laboratory, hosted by University of Nice - Sophia Antipolis). This is a platform for lightweight components for service composition SLCA (Service Lightweight Component Architecture) which enables the design of ambient computing applications by assembling software components, orchestrating access to services through infrastructure devices from ambient. WComp supports protocols such as UPnP (Universal Plug and Play) [37] and Web services, allowing components through the proxy to interact with them.

WComp is a prototyping "development" environment for context-aware applications. The WComp Architecture is organized around Containers and Designers paradigms. The purpose of the Containers is to take into account system services required by Components of an assembly during runtime: instantiation, destruction of software Components and Connectors. The purpose of the Designers allows configuring assemblies of through Containers. To promote adaptation to context WComp uses Aspect [27] Assembly paradigm. Aspect Assemblies can either be selected by a user or fired by a context adaptation process. It uses a weaver that allows adding and or suppressing components. A container includes a set of (Beans) components and each bean has: properties, input methods that use received input information, and output Methods to send to another bean, for instance, output information.

Aspect Assemblies allow defining links between Beans by using input and output information. WComp uses UPnP (Plug and Play) technology to detect locally whether the device is active or not and to define input methods and sent events for each component. With this architecture WComp allows: i) managing devices heterogeneity and dynamic discovering by using UPnP, ii) events driven interactions with devices, iii) managing dynamic devices connection and disconnection (dynamic re configuration on run time) in infrastructure. Let us see now the proposed solution.

## 4. TOWARDS SELF ADAPTABILITY:
### 4.1. Needs of self adaptability

The adaptation of service-oriented architectures is at the heart of building new applications. A self-adaptable architecture is that it takes into account the business logic and HMI (Human Machine Interface). A workflow engine allows you to run a process defined elsewhere in the tool design process that accompanies it. This execution is sequential and is completely unconditional, conditional and is based on a set of rules defining the conditions of connection. The workflow tools allow the definition of rules more or less sophisticated but still generally quite simple and few: Boolean operators, data fields of the process values entered by operators, properties of any documents involved in the process, etc..

The workflow engine can then connect to the rules engine to "know" what option to take during the course of a process. The rules engine also allows users to expose simple interfaces for generating these rules.

Following the research and the state of the art described above, lead us to determine the list of needs:
- An SOA (Service Oriented Architecture) [5] for its flexibility, interoperability with Web services.
- Use of aspects for adaptation to changes that do not provide Web services.
- The use of such powerful concepts: reflexivity and / or MOP (Meta Object Protocol) [24] to gain flexibility and performance, or dynamic reconfiguration.
- Adaptation to the prevailing situation with mechanisms and platforms to capture the events and process them in real time, or in a near real time, by invoking the right service.
- Consideration of transverse functions such as security, log management, ...
- Consideration of management rules intrinsically.
- The management of processes and data.
- Interoperable intelligent connectors that can be applied to the platform regardless of its technology.

### 4.2. Feasibility
#### 4.2.1. Framework

We chose the platform WComp particular interest to us for several reasons:
- Respect for the black box modules (devices ambient space which we have no access except through their interfaces or boundaries software / hardware). This is an implementation deliberately masked by patents, for example.
- The event communication to effectively manage the intrinsic reactivity of ambient computing systems.

- Modularity is at its peak transverse to manage the adaptations: creation of an entire application from producing aspects of structural reconfigurations rather than code injection to meet the first point: the adaptation to the emergence.

WComp represents a graphical interface for dynamic reconfiguration of components. This is an interface used to edit BSL (BioSafety Level) or ISL4WComp (Interaction Specification Language for WComp). The concept of aspect assembly (AA) consists of four elements: The first, represents the "graft", is an editor for editing the specification of a graft. The second is the "cut point", which allows registering the points of cuts. Finally, the last two elements are used to specify the name of the AA and then select view AAs applied to the assembly.
BSL and the ISL4WComp designer aim to select the AAs and weave them to modify the target assembly Composite Service.

WComp is an assembly of components encapsulated in a composite service. This approach relies on local assemblies and components communicating by light waves of events. Set points and propose a solution for adaptable applications in the ambient computing. This is an approach to manage the adaptivity of applications by the notion of aspect assembly.
A Component WComp (Figure 1) is a software component characterized by properties, methods of entry using the input information and methods that the events issues that send information to other software components to provide a service according the request of a customer.

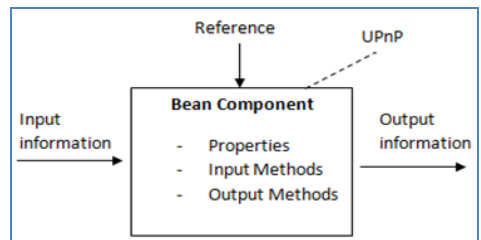

Figure 1. Beans in WComp [26]

A component is created by:

- The developer defines a class in C # characterizing the component.
- The detection of a device or an external machine using UPnP technology. This will be the representative of the device that requires input methods and events defined by the UPnP automatically.
- The WSDL proxy represents a remote service.

Each component uses system resources or WComp. These resources must be referenced in the component through importing files which represent system resources.

We distinguish two levels of manage application WComp depending on the appearance and disappearance of devices:

- Local level: represents an approach to composition LCA (Lightweight Components Architecture) that allows us to define the components as needed.
- The level distribution: SLCA approach to composition that communicates with other devices using UPnP technology.

In Figure 2, the links: (1), (2), (3) and (4) represent the information entered by the user or the client to the container which contains the various components for use in an application under WComp. The link (5) represents the information generated. This information is processed by the events emitted in the component. The link (6) represents the level where business is done between different assembly components. The links (7), (8) and (9) represent the level integration. In this level, WComp uses different technologies to dynamically and automatically detect devices to provide a service according to customer demand or context. WComp offers a tool to invoke remote Web services; this tool is called "WSDL proxy" to define a path that represents a WSDL Web service. WComp also offers another tool called "UPnP proxy" to detect the devices based on Microsoft UPnP.

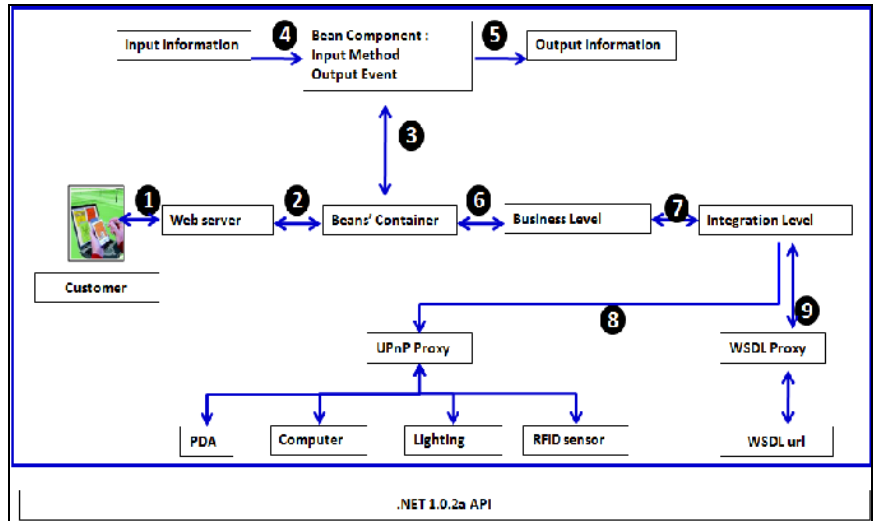

Figure 2. Implementation Scenario in WComp.

Based on this architecture and the component representation, allows WComp:
- The dynamic discovery of devices using UPnP technology.
- The interaction events with the devices.
- Manage the appearance / disappearance of dynamic devices in the infrastructure.

This platform offers several functionality:

- Composite service model

WComp involves the introduction of the notion of composite service which is to implement a distributed service by assembling software components localized light.

The analysis of communication relies on WComp communication events in composite services for ambient spaces.

- Component model

WComp offers specific components composite service. The probes and wells used to construct a composite service interface provided to and exports the data contained in the assembly. This allows exporting features to keep the internal capacity of reusability. Finally, to manage the reactivity, the interfaces of the components consist of a set of entry points and events.

- Adaptivity transverse model

The adaptivity should take into account the highly dynamic ambient space. Therefore, it must provide a software reconfiguration mechanism that reacts to every change in the ambient space.

The platform WComp is installed under the 1.0.2a SharpDevelop [38] version in the .Net1.1 framework. Nous avons essayé d'implémenter un moteur de règles avec cette version mais nous avons constaté qu'elle n'est pas performante. C'est pour cette raison et pour montrer la faisabilité de notre approche, nous avons choisi la version 3.2 de Sharpdevelop sous l'environnement de travail .Net 3.5 qui est plus évoluée et elle intègre les flux.

Under the 3.2 SharpDevelop version, we used WWF which is both the programming model, the engine and tools to quickly generate applications that support the flow on Windows.
Under WComp we have integrated a rule engine that can provide management rules that deal with business logic. The rules engine can communicate with a workflow engine, which helps optimize and evolution of these assemblies separating the events produced by the components defined in an application and WComp.

This integration solves simple scenarios such as displaying user interface controls based on user input or complex scenarios encountered by large enterprises.

Our solution is based on the WWF under. Net that can solve simple scenarios such as displaying user interface controls based on user input or complex scenarios encountered by large enterprises.

### 4.2.2. Functional architecture WComp with WWF

Figure 3 shows the functional architecture of our system. In this architecture we introduced the functions that the information system must support and internal organization in relation to the platform WComp.

The various functionalities offered by this system are the security and administration in treating the business logic from the workflow and rules. The invocation of distant and ambient services is also permitted by this architecture using technologies dedicated to each type of invocation.

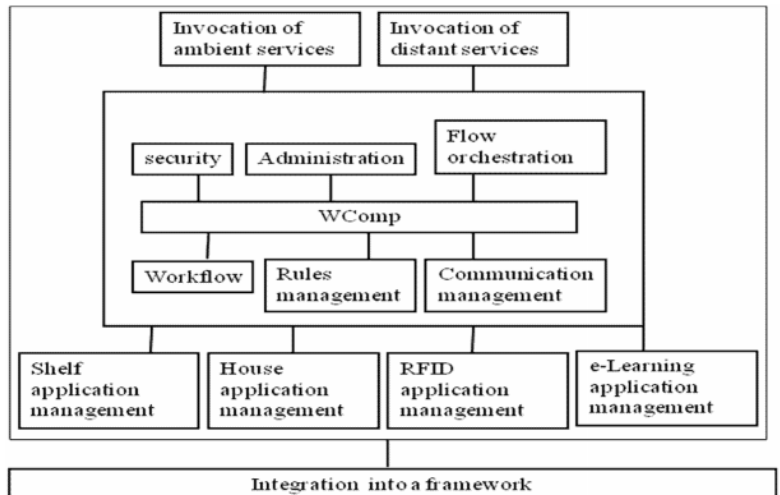

Figure 3. Functional architecture WComp_Workflow

### 4.2.3. Technical Architecture WComp with WWF

This architecture (Figure 4) allows the structuring of technical capabilities and infrastructure in our new approach Wcomp.

In this architecture, except for the different needs initially used by Wcomp (service invocation ambient and remote data orchestration ...), we integrated connectors rules engine that communicates with a workflow engine in framework .NET. In this rules engine we need to define the rules that manage the data flow to finally produce events providing services to the customer. The information shall be provided from a component that specifies the service to send it to another component by assembling them in a container through the language of Aspect of Assembly.

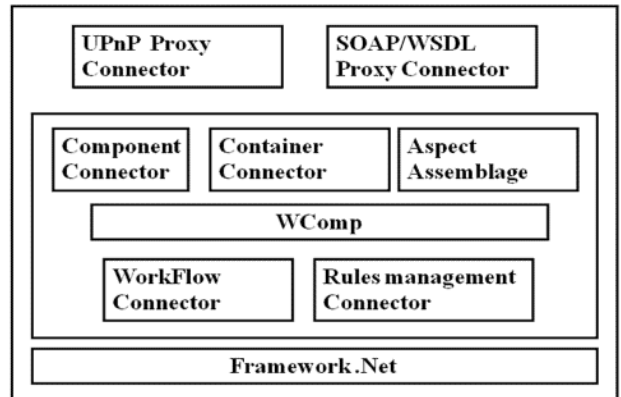

Figure 4. Technical Architecture WComp_Workflow

## 4.3. Case Study
### 4.3.1. Description

We chose to take a sample case study of authentication. This authentication is supposed to capture with a RFID (Radio Frequency Identification) [42] [18] sensor using UPnP technology in WComp. This authentication can, thereafter, to monitor such access to a well determined. In our case, and to simulate this case study, we created a man-machine interface that captures a user authentication, this authentication is then verified based on business rules defined in XML, to finally show a message validation or invalidation of the value set as authentication.

### 4.3.2. Implementation

In a first step, we built the container WComp under sharpdevelop1.0.2a. This container contains all components necessary to make the bean test authentication with an assembly between them. In a second step, we imported the container WComp sharpdevelop3.2 under a project to integrate the flows and rules. We chose to show the code of the rule that deals with the flow and data entered by the user. This code is in XML form.

```
15   <ns0:CodeBinaryOperatorExpression.Right>
16   <ns0:CodePrimitiveExpression>
17   <ns0:CodePrimitiveExpression.Value>
18   <ns1:String xmlns:ns1="clr-namespace:System;Assembly=mscorlib, Version=2.0.0.0, Culture=neutral,
19   PublicKeyToken=b77a5c561934e089">Felhi</ns1:String>
20   </ns0:CodePrimitiveExpression.Value>
21   </ns0:CodePrimitiveExpression>
22   </ns0:CodeBinaryOperatorExpression.Right>
23   </ns0:CodeBinaryOperatorExpression>
24   </RuleExpressionCondition.Expression>
25   </RuleExpressionCondition>
26   </RuleDefinitions.Conditions>
27   </RuleDefinitions>
```

**Code 1 : Setting a value.**

In this block we defined as an example the rule that gives the exact value of authenticating a user. As shown in line 19, the exact value is "Felhi". This value is normally detected by RFID and simulated in this example by entering a "TexField", and for example displayed on a screen that is simulated by a "label".

## 5. CONCLUSIONS AND FUTURE WORKS

For our needs to adaptability, we studied the platforms by exposing their operating principle. And in particular we studied the platforms for adaptabilities context, CORTEX and WComp. This study aims to initiate our future research to provide web services self-adaptability to context in SOA.

We presented a proposal to an self-adaptable SOA. We've built a rules engine within WComp which can offer management rules that deal with business logic. Business logic can help in the development and optimization of these assemblies separating the events produced by the components defined in an WComp application.

This integration is simplistic since we looked at the feasibility of our solution using connectors available at the .NET platform. We focused our work on the inclusion of rules on management WComp. We hope to extend this business rules engine to transverse rules such as: security, administration, etc... This solution is a first step towards our problem of defining a service-oriented middleware (self) adaptable to the context.

## ACKNOWLEDGEMENTS
We thank everyone!

**Authors**

**Faîçal Felhi** received the master degree in Intelligent Information System in 2010. He is currently a PhD student in the BESTMOD laboratory in the high institute of management of Tunis. - Tunisia. He is actually teaching in the high institute of Computer and Mathematic of Monastir - Tunisia.

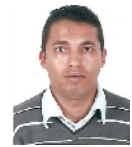